\documentclass[12pt]{article}
\usepackage{amsfonts,latexsym}

\title{Light Sheets\\ and the\\ Covariant Entropy Conjecture}
\author{Robert J Low\thanks{email: r.low@coventry.ac.uk} \\
        Mathematics Dept,
        Coventry University,
        Coventry CV1 5FB,
        UK}

\begin{document}

\maketitle

\begin{abstract}
We examine the holography bound suggested by Bousso in his
covariant entropy conjecture, and
argue that it is violated because his notion of light sheet
is too generous. We suggest its replacement by a weaker bound.
\end{abstract}

\section{Introduction}
Since the original papers of 't Hooft and Susskind
\cite{thooft93,suss95}, much work has
been done on the notion of holography; see Smolin \cite{smolin01}
and references therein for a review. In particular, 
Fischler and Susskind \cite{fs98} suggested a holography
bound in terms of null surfaces in the context of 
cosmological solutions. Bousso
\cite{raph991,raph991,raph00} has since stated and made 
use of a conjectured holographic entropy bound which 
generalises the work  of Fischler and Susskind by relating 
the area of a two-surface to the entropy contained on a section 
of null hypersurface which ends on that two-surface
in general space-times. This work also 
builds on earlier work of Bekenstein \cite{bek81} relating 
the entropy of a region to its surface area under certain 
constraints.

We briefly recall the Fischler-Susskind-Bousso (FSB) bound.

Let $\Sigma$ be a spacelike two-surface in a space-time, $\cal M$.
We wish to give a bound in terms of $A$, the area of $\Sigma$, for
the entropy contained within $\Sigma$. There is an ambiguity here, 
since many three-surfaces in space-time span the surface $\Sigma$.
One approach to removing this ambiguity is to
obtain a natural candidate for the region bounded by $\Sigma$ whose
entropy is to be considered. So consider the four orthogonal null 
(geodesic) congruences to $\Sigma$; then a light sheet, $L$, is 
given by an orthogonal null congruence which has negative expansion 
everywhere on $\Sigma$, each null geodesic being extended to the 
first point at which the expansion vanishes (i.e. the first point 
along the null geodesic which is conjugate to $\Sigma$). Bousso's 
conjecture is, then, that the entropy
contained on $L$, denoted by $S(L)$, satisfies the inequality
$$
S(L) \leq \frac{A}{4}.
$$
This inequality is the FSB bound, and the statement that it holds
in general space-times is the covariant entropy conjecture.
It is supported by various examples in the context
of standard cosmological solutions. However, the explicit examples
provided have relied heavily on $\Sigma$ having spherical symmetry.
It is the aim of this paper to argue that in the more general
situation where spherical symmetry of $\Sigma$ is dropped, the
FSB bound is violated, and that in general, the
light sheet should only include that portion of the null congruence
up to the first conjugate point or point of self-intersection on each null
geodesic.

It should be noted that this argument does not attempt to invalidate
the programme being carried out by Bousso, only to establish that
the inequalities used are too strong, and should be replaced by
weaker ones when appropriate.

\section{Ellipsoidal examples}
The motivation for the following section is the observation
that in Bousso's examination of the covariant entropy
conjecture, situations were noted where the inequality was
saturated, even in the case where $\Sigma$ has spherical
symmetry. For such surfaces, the light sheet is
cut off by conjugate points before any self-intersection
occurs. This suggests that it might be possible to find
a counter-example by considering a light sheet at least part
of which extends far beyond the first self-intersection
before the first conjugate point occurs.

To this end,
we will consider first the situation for a field of matter
on a Minkowskian background, and then extend this to the
consideration of full solutions of the Einstein equations.

So we begin by considering the ellipsoid, $\Sigma$, given by
$$
\frac{x^2}{a^2} + \frac{y^2}{a^2} + \frac{z^2}{b^2} = 1
$$
in the $\{t=0\}$ slice of Minkowski space, with the
usual coordinates $(t,x,y,z)$. Then take as light sheet, $L$,
the relevant portion of the future pointing inward directed
orthogonal null congruence to $\Sigma$. It is clear than
as $b$ approaches 0, the ellipsoid approaches (two copies of)
the disk $x^2 + y^2 \leq a^2$, while the associated light
sheet approaches two pencils of null geodesics.

By taking $b$ sufficiently small, the area, $A$, of $\Sigma$
approaches $2\pi a^2$ arbitrarily closely, while the extent
of $L$ increases without bound. It then follows that if Minkowski
space is the background on which lies a stationary matter field whose
entropy density is constant, then by taking
$b$ sufficiently small, the entropy passing through $L$ may
be made unboundedly large, and so for some $b>0$, the FSB
bound must be violated.

This tells us that the covariant entropy conjecture 
fails for stationary matter on a flat background (which 
is a first approximation to the cosmological case), but 
of course this does not necessarily imply that the 
conjecture itself must be violated; it merely
suggests that concern is justified. We proceed in
steps of decreasing idealisation.

To this end, we next consider a FLRW space-time with constant
co-moving entropy density, as in Tavakol and Ellis \cite{tavell99}.
In this case, we consider an ellipsoid in a surface of
constant time. By taking the ellipsoid sufficiently small,
we obtain a situation which is well-modelled by flat space-time
again. Thus, by taking $a$ and $b$ sufficiently small,
the extent of  the light sheet can be made large compared to 
the area of $\Sigma$, and once more we  expect that for sufficiently  
small $b$, the entropy contained on  the light sheet, $S(L)$, will 
become too great to be bounded by $A/4$.

As a next step, we consider a general (no longer assumed to be
conformally flat)  space-time, in which
the entropy is modelled by a density bounded away from zero
and non-decreasing along any future directed timelike curve.
Once more, by restricting our attention to a sufficiently
small region of space-time, we can assume that the deviation
from a flat metric is arbitrarily small, and the conclusion
will once again follow.

Thus we can conclude that in the case where entropy is
modelled as a density on some fixed background geometry,
the FSB bound is violated.

Each of these cases is suggestive, but in order to
argue that the conjecture itself fails, the situation 
must be adapted to a full solution of the Einstein equations.
To this end, we should note that the above argument takes 
place in a homogeneous background, as does much of the discussion
of Bekenstein, Fischler and Susskind, and Bousso, cited above.
However, in a real space-time there are fluctuations and
inhomogeneities, whose presence must be acknowledged, and indeed 
these inhomogeneities are essential to the discussion of
entropy.

Note that the effect of such fluctuations provides the
basis of the arguments of Tavakol and Ellis \cite{tavell99}
that the light sheet should be truncated at self-intersections.
Their argument, though, is based on the observation that truncating
only at conjugate points yields a surface of extremely complicated
geometry, rendering the light-sheet effectively uncomputable. They
suggest truncating earlier as a cure for this problem of operational
definability. It is the intent of the current paper to strengthen
their arguments by arguing that the covariant entropy conjecture
is in fact violated, not just difficult to work with.

Fortunately, the presence of such fluctuations 
does not destroy the argument presented in the previous section. For
if the size of the fluctuations in the energy density of the
matter fields is small compared with the energy density, then
we expect only a small proportional change in the distance along
each null geodesic to the first point where the expansion vanishes.
(Even though the locus defined by such points becomes much more
complicated.)

We thus expect that just as in the case where entropy is
modelled as a density on an averaged background, the amount
of entropy on the light sheet will become large compared
to the area of our small spacelike ellipsoid, if it is
sufficiently oblate.
Therefore we expect the covariant entropy conjecture to be
false in general for such sufficiently small oblate ellipsoids.
We conclude from that above considerations that the light
sheet used in the holography bound is too large, and that
a weakened form of the covariant entropy conjecture would
be more useful.

\section{Discussion}
We should now address the issue of just how this suggests that we 
weaken the covariant entropy conjecture. A motivation is provided
by a standard result from causal theory. 

Suppose, then, that $\Sigma$ is a spacelike two-surface lying
in a Cauchy surface, $\cal S$, of an open globally hyperbolic space-time 
$\cal M$ (so that the notion of inside $\Sigma$ is unambigous), 
such that the inward directed, orthogonal null congruence $\cal C$
composed of null geodesics proceeding to the future has negative 
expansion everywhere. Let $L$ be the light sheet associated with 
$\cal C$, and let $B$ denote a spacelike three-surface with
boundary $\Sigma$. ($B$ is not assumed to lie in the Cauchy
surface mentioned above.)

Now, consider a point $p \in L$ which lies
to the future of a point of self-intersection, $s$, of $\cal C$. 
As Bousso and Tavakol and Ellis observe \cite{raph991,tavell99} this may 
lead to a multiple counting of some of the entropy passing through 
$B$; thus $S(L)$ gives an upper bound for the entropy of any
spacelike three-surface spanning $\Sigma$. But the situation is
rather worse than this observation alone indicates.

For $q$ is connected to $s$ by a future-directed
null geodesic segment, $\gamma_1$, and $s$ is connected to $p$ by
a future-directed null geodesic segment $\gamma_2$, where $\gamma_1$
and $\gamma_2$ are not segments of a single null geodesic. It
follows \cite{HE73} that $q \in I^-(p)$. 
But since $I^-(p)$ is open, and $p \in \partial B$, it follows that there
are points in $\cal S$ outside $B$ to the chronological past of $p$. 

This implies that there are points outside $D(B)$, the Cauchy 
development of $B$, which lie to the past of $p$. As a consequence, 
entropy from regions which
cannot be regarded as inside $\Sigma$ can contribute to the entropy
on the light sheet. Indeed, in the case of the (nearly) flattened
ellipsoid described above, most of the entropy on the light sheet
comes from a large region outside $\Sigma$. Thus entropy passing
through $L$ may greatly exceed that contained in any spacelike 
three-surface spanning $\Sigma$.

This provides a physical interpretation for the
violation of the FSB bound, and strongly suggests
that we should also stop the light sheet at points of self-intersection,
thereby ensuring that inappropriate entropy does not contribute.
Denote this truncated light sheet by $L'$.

In the case where the future horismos of $B$, $E^+(B)$, is composed of
null geodesic segments starting on $\Sigma$, and this congruence
has negative expansion everywhere on $\Sigma$, $E^+(B)$ coincides
with $L'$. We then see that in this case the
entropy passing through the light sheet all starts off at some point
inside $\Sigma$, and there is no contribution from regions outside
$\Sigma$. Thus in this case, the entropy contained on the (truncated)
light sheet, $S(L')$ satisfies $S(L) \geq S(L') \geq S(B)$.
The first inequality follows from the fact that $L' \subseteq L$,
and the second from the second law of thermodynamics.
Indeed, $S(L')$ is the least upper bound of $S(B)$ where $B$ ranges
over all three-surfaces spanning $\Sigma$. This seems to be
a suitable choice of light sheet, and again allows the recovery
of the Bekenstein bound in the relevant situations.

\section{Conclusion}
We have argued that the covariant entropy conjecture is violated
for sufficiently oblate small ellipsoidal surfaces, and provided
a physical interpretation for the source of the violation. This
suggests a particular adaptation of the light sheet in which we
do not include points beyond self-intersections of the null congruence.
Finally, we note again that this does not attempt to invalidate the
holographic entropy programme; it rather argues that some of the
inequalities used so far need to be replaced by rather weaker
ones.

\noindent
\textbf{Acknowledgement} It is a pleasure to thank Andrew Chamblin
of the Centre for Theoretical Physics, MIT,
for introducing me to the covariant entropy conjecture, 
subsequent interesting discussions, and helpful comments on
the manuscript.


\thebibliography{99}
\bibitem{thooft93} G. 't Hooft (1993) 
\textit{Dimensional reduction in quantum gravity}
in \textit{Salamfetschrift} ed A. Alo,. J. Ellis,
S. Randjbar-Daemi (World Scientific); gr-qc/9310026.

\bibitem{suss95} L. Susskind (1995) \textit{The world as a hologram}, 
J. Math. Phys. \textbf{36}, 6377--6396; hep-th/9409089.

\bibitem{smolin01} L. Smolin (2001),
\textit{The strong and weak holographic principles}
Nucl. Phys. \textbf{B601}, 209--247;
hep-th/0003056.

\bibitem{fs98} W. Fischler \& L. Susskind (1998)
\title{Holography and cosmology}; hep-th/9806039.

\bibitem{raph991} R. Bousso (1999)
\textit{A covariant entropy conjecture} 
JHEP \textbf{9907} U58--U91; hep-th/9905177.

\bibitem{raph992} R. Bousso (1999) 
\title{Holography in general space-times}
JHEP \textbf{9906} U534--U556; hep-th/9906022.


\bibitem{raph00} R. Bousso (2000)
\textit{The holographic principle for general backgrounds}
Class. Quant. Grav. \textbf{17} 997--1005;
hep-th/9911002.

\bibitem{bek81} J. D. Bekenstein (1981)
\textit{A universal upper bound on the entropy to energy
ratio for bounded systems} Phys. Rev. D 
\textbf{23} 287--298

\bibitem{tavell99} R. Tavakol \& G. F. R. Ellis (1999)
\textit{On holography and cosmology}
Phys.Lett. \textbf{B469}  37--45 ; hep-th/9908093

\bibitem{HE73} S. W. Hawking \& G. F. R. Ellis (1973)
\textit{The large scale structure of space-time}
Cambridge University Press
\end{document}